\documentclass{an}
\usepackage{graphicx}
\usepackage{times}
\usepackage{fancyhdr}
\usepackage{psfig}
\newcommand{\HI}{{\ion{H}{1}}}

\newcommand{\abHI}{\rm H{\hskip 0.05cm\scriptsize I}}

\newcommand{\kms}{$\,$km$\,$s$^{-1}$}

\newcommand{\mJybeam}{mJy beam$^{-1}$}

\def\HI{H{\,\small I}}

\def\OIII{[O{\,\small III}]}

\begin{document}

\title{Neutral hydrogen in radio galaxies: results from nearby, 
importance for far away
}

\author{Raffaella Morganti}
\institute{Netherlands Foundation for Research in Astronomy, Postbus 2,
NL-7990 AA, Dwingeloo, The Netherlands \\
Kapteyn Astronomical Institute, University of Groningen
P.O. Box 800, 9700 AV Groningen, The Netherlands
}

\date{Received; accepted; published online}

\abstract{The study of neutral hydrogen emission and absorption  in radio
galaxies is giving new and important insights on a variety of phenomena
observed in these objects.  Such observations are helping to understand the
origin of the host galaxy, the effects of the interaction between the radio
jet and the ISM, the presence of fast gaseous outflows as well as jet-induced
star formation.  Recent results obtained on these phenomena are summarized in
this review.  Although the \HI\ observations concentrate on nearby radio
galaxies, the results also have relevance for the high-$z$ objects as all
these phenomena are important, and likely even more common, in high-redshift
radio sources.
}
\correspondence{morganti@astron.nl}

\maketitle

\section{Introduction: what can we learn from the \HI\ in nearby radio galaxies
}

There are a number of important aspects of radio galaxies that can be
investigated using observations of neutral hydrogen. Because of sensitivity
limitations of present day radio telescopes, some of these studies are restricted
to nearby radio galaxies. Nevertheless, they can give important insights on
phenomena that are likely to be very common in high redshift objects.
Here I summarize some recent results obtained for low-$z$ radio galaxies,
underlining their relevance for galaxies in the far away Universe.

\begin{figure*}
\centerline{\psfig{figure=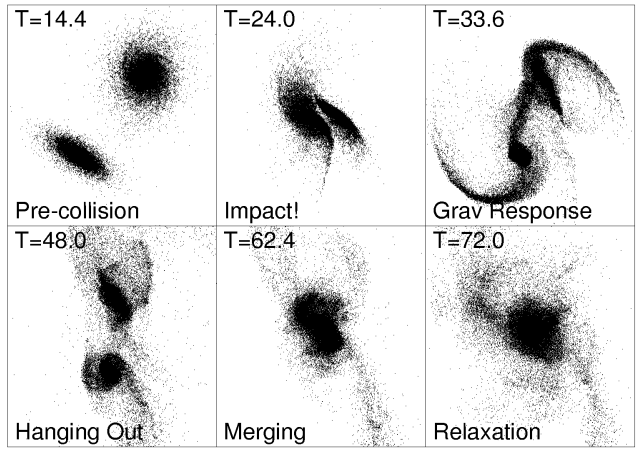,width=9cm,angle=0}}
\caption{Numerical simulations of major mergers resulting in an
elliptical galaxy (from Mihos 1999).
 }
\end{figure*}

Hierarchical (major) merging and accretion of small clumps appears to be a
good description of the formation of early-type galaxies, i.e. the typical
host of radio galaxies (see Fig. 1, Mihos 1999).  If this is the case, the
presence and morphology of extended \HI\ can be used as tracer to investigate
the origin and evolution of these galaxies.  This is particularly interesting
in the case of radio galaxies as the origin of activity in galaxies is often
explained as triggered by merger and/or interaction processes.  Torques and
shocks during the merger can remove angular momentum from the gas in the
merging galaxies and this provides injection of substantial amounts of
gas/dust into the central nuclear regions (see e.g.\ Mihos \& Hernquist
1996). Indeed, this appears to be the case for radio galaxies as suggested by
morphological and kinematical evidence (e.g.\ Heckman et al.\ 1986; Tadhunter
et al.\ 1989).

Another aspect that can be investigated using the neutral hydrogen is the
effect of the interaction between the radio plasma and the interstellar
medium (ISM) that surrounds the radio source.  Indeed, the phase of nuclear activity
is now increasingly recognized to play an important role in the evolution of
the galaxy itself.  Particularly important in this respect are gas outflows
that can be generated by this activity and the effect they can have on the
ISM.  This feedback can be extremely important for the
evolution of the galaxy, up to the point that it could limit the growth of the
nuclear black-hole (e.g.\ Silk \& Rees 1998; Wyithe \& Loeb 2003).  Thus, the
processes of assembly of the host galaxy, the supply of gas to the central
region, as well as the effects that the triggering of the (radio) activity has
on this gas, are tightly related and essential for our understanding of radio
galaxies.  An other aspect related to the interaction is the possibility of
jet-induced star formation, a phenomenon considered to be particularly
relevant for high-$z$ radio galaxies.

\begin{figure*}
\hfill
\includegraphics[width=.4\textwidth]{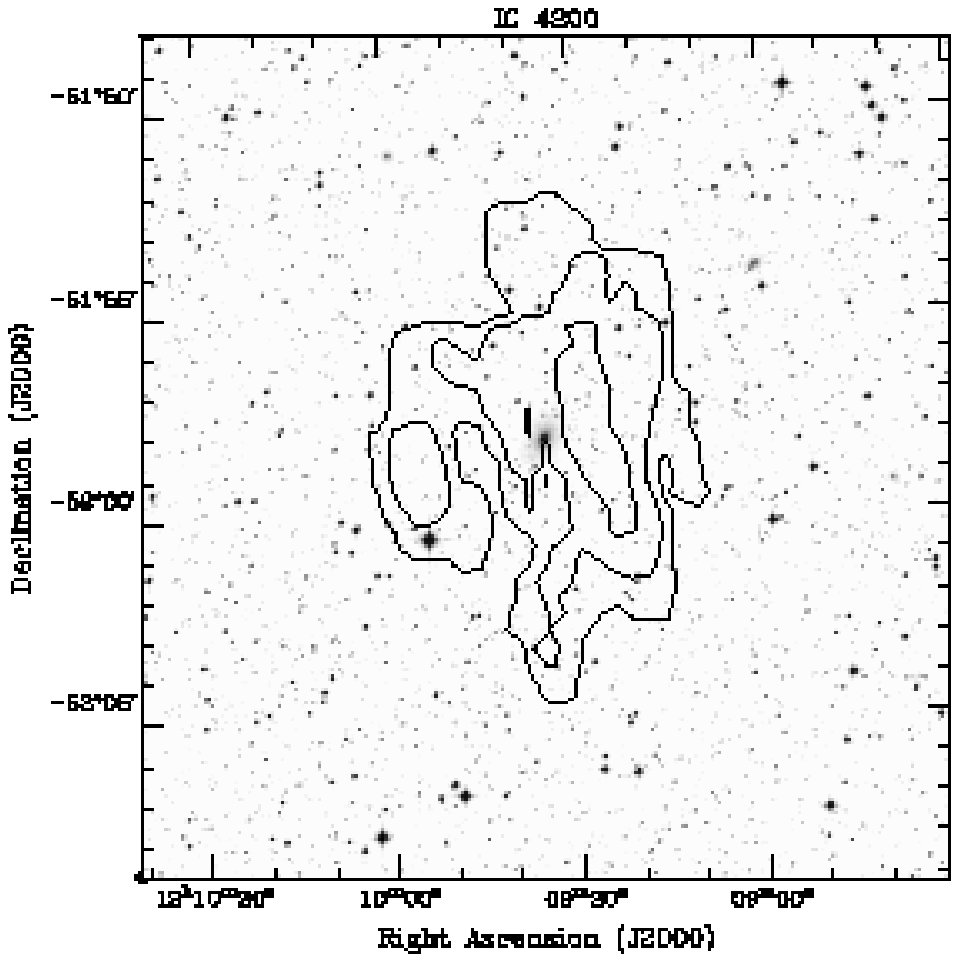}
\hfill
\includegraphics[width=.4\textwidth]{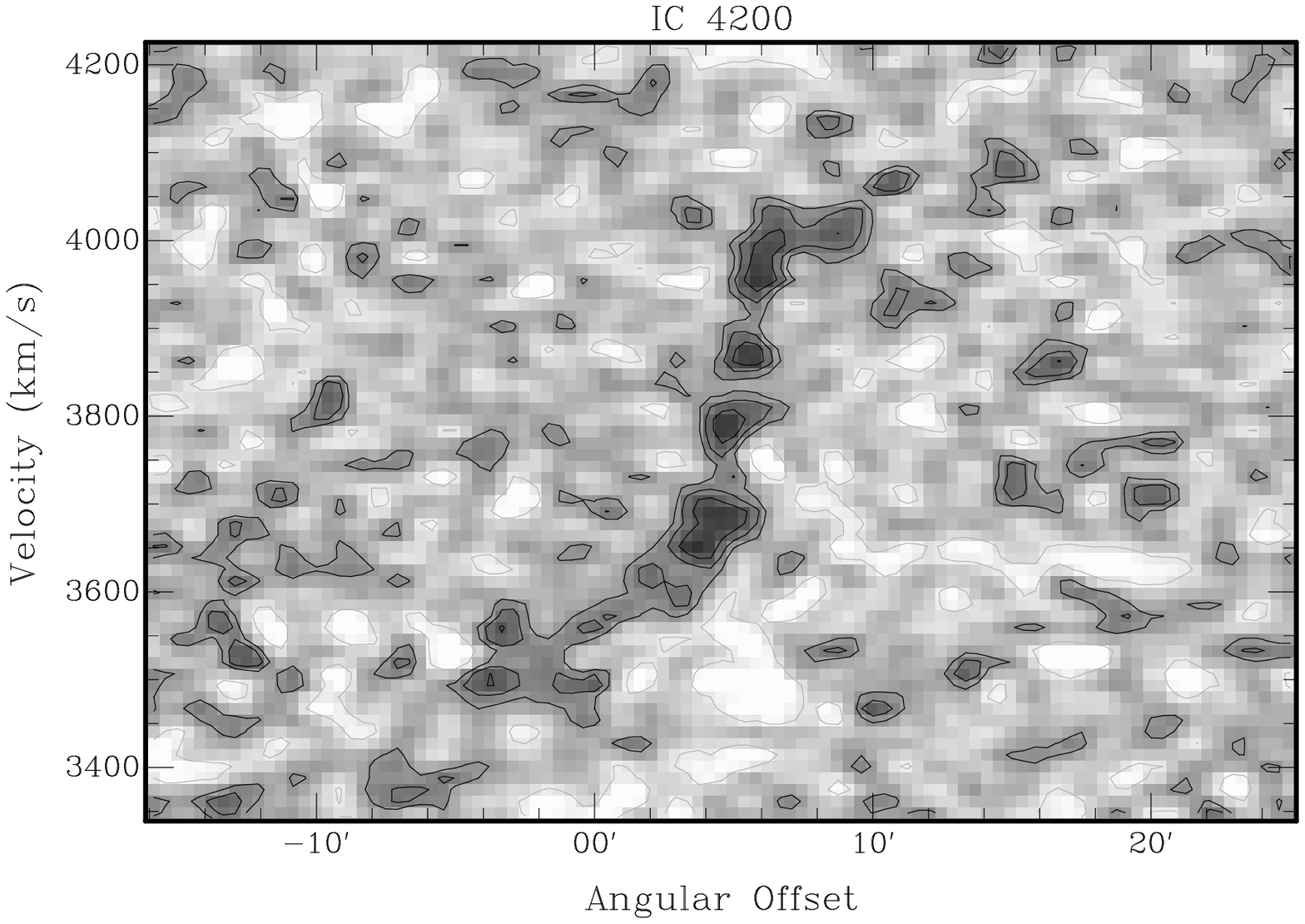}
\hfill\null
\caption[]{Total intensity (left) and position-velocity plot along the
major axis (right) for the early-type galaxy IC~4200.  The
\HI\ disk is about 200 kpc in size.  Contour levels are:
$2, 4, 8, 16 \times 10^{19}$ cm$^{-2}$ (left)
and  -3, 3, 6, 8 mJy beam$^{-1}$ (right).  }
\end{figure*}

\section{Large-scale \HI\ (emission) in nearby early-type galaxies}

Large-scale \HI\ (in emission) around galaxies hosting radio sources provides
an important signature of whether a (major) merger has occurred in the life of
these galaxies.  More details about the results of this study are presented in
Emonts et al. (these Proceedings). However, to put these results in the more
general context of the formation and evolution of early-type galaxies, we
describe first the results obtained on the occurrence and morphology of the
neutral hydrogen in "normal" early-type galaxies.  These results will be later
(Sec 2.3) compared with what found for radio galaxies.

\subsection{Shallow surveys}

It is already known since many years that HI-rich early-type galaxies do exist
(e.g.
Knapp et al. 1985; Morganti et al. 1997; van Gorkom etal. 1997; Oosterloo et
al. 2002).  More recently, however, a systematic, albeit shallow, \HI\ survey
of all early-type galaxies south of $\delta < -25^{\circ}$ with $V < 6000$ km
s$^{-1}$ (based on the Parkes All Sky \HI\ Survey: HIPASS, Barnes et al. 2001
and followed up with the ATCA, Oosterloo et al. in preparation) has revealed that in
most of these \HI-rich early-type galaxies the \HI\ is distributed in very
large (up to 200 kpc in size), regular disks of low column density
\HI.  Fig.  2 shows the case of IC~4200.  {\sl The amount of \HI\ and
the size of the structures can be explained as result of major mergers between
gas-rich disk galaxies}.  This is indeed how at least some early-type galaxies
are believed to be formed (see e.g. the case of NGC~7252, Hibbard \& van
Gorkom 1996).

As mentioned above, this survey is shallow, therefore able to detected only
galaxies with associated more than $10^9$ $M_{\odot}$ of neutral
hydrogen. Nevertheless, the results are telling us not only that such large
amount of neutral hydrogen (with \HI\ masses up to $10^{10}$ $M_{\odot}$ and
log M$_{\rm \HI}$/L$_B$ is between -1 and 0 for the \HI\ detected galaxies)
appear to be present around 5-10\% of early-type galaxies but also that,
because of their regular and large appearance, these disks must be quite old,
in many cases well over $5 \times 10^9$ yr. Hence, they are not related to
recent accretions.  The \HI\ column density in these disks peaks at only
$10^{20}$ atoms cm$^{-2}$ (i.e. 0.5-1 M$\odot$/pc$^2$). Thus, despite the
large \HI\ reservoir, no significant star formation is occurring, therefore
the HI is not used and the \HI\ disks will evolve only very slowly.  Finally,
these large, regular structures can be used to get information about the dark
matter content of early-type galaxies. So far it appears that this is similar
to what found in spirals galaxies (see Fig. 3).

\begin{figure}
\centerline{\psfig{figure=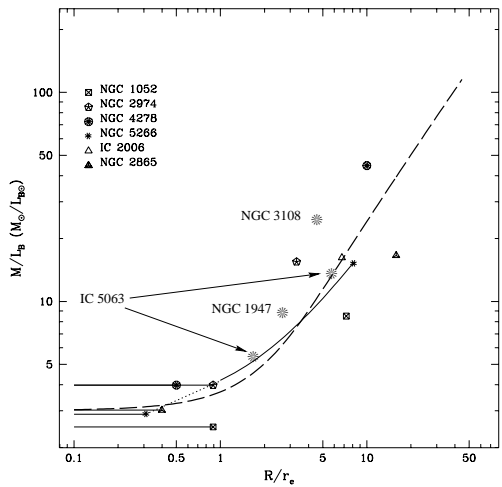,width=7cm,angle=0}}
\caption{ The log($M/L_B$) - log($R/R_{\rm e}$)
diagram (modified from Bertola et al.\ 1993).  The dashed thick line
represents the cumulative $M/L_B$ as function of radius for spiral galaxies.
The symbols represent the data for the elliptical galaxies obtained from the optical
and the \HI\ data. From Morganti et al. 1999.}
\end{figure}

\subsection{Deep \HI-search}

The shallow surveys described above have shown that \HI-rich early-type galaxies
do exist. However, it is also important to investigate how common is, around
these galaxies, the presence of (even a modest amount of) neutral hydrogen and
what are its characteristics. To investigate this, a representative sample of
12 early-type galaxies has been observed using the WSRT. These galaxies were
selected because they are part of the sample studied in the optical with the SAURON
panoramic integral-field spectrograph on the William Herschel Telescope, La
Palma. SAURON provides a very detailed view of the kinematics of the ionised
gas and of the stellar component in the inner regions of these objects (de
Zeeuw et al. 2002).

This deep, albeit so far small, survey can detect \HI\ masses down to
$<10^7$ M$_\odot$. In nine of the 12 objects observed (75\%) \HI\ associated
with the galaxy has been detected (Morganti, de Zeeuw, Oosterloo et
al. 2005a). This finding indicates that {\sl neutral hydrogen appears to be a
common characteristic of these galaxies, provided that deep enough
observations are available}.  The neutral hydrogen shows a variety of
morphologies, from complex structures (tails, offset clouds) to regularly
rotating disks. Four of the detections belong to the latter group and they
will be the main test cases for probing the dark matter content in the inner
and outer part of the galaxies in a consistent way. The most extreme case in
our sample is a gas disk of only $2 \times 10^6$ M$_\odot$ and a column density
of $5 \times 10^{18}$ cm$^{-2}$ that has been found in the young SO galaxy
NGC~4150.

Figure 4 shows the results for the elliptical galaxy NGC 4278. This galaxy has
an \HI\ disk that extends well beyond the optical image. The velocity maps of
the ionised gas and of the stars (as obtained with SAURON) together with the
velocity map of the \HI\ illustrate one of the recurrently found
characteristics: the kinematics of the \HI\ is very similar to that of the
ionised gas, indicating that they form one single structure. This is
underlined by the fact that galaxies with little or no ionised gas are less
likely to show \HI.  In NGC~4278, an offset exists between the position angle
of the rotating gas disk (\HI\ and ionised) and that of the rotation of the
stars, something that is observed in several cases and may reflect the
non-axisymmetric nature of the galaxies. These offsets in kinematic alignment
can vary from co- to counter-rotating, sometimes showing dramatic twists
within a single object.

\begin{figure}
\centerline{\psfig{figure=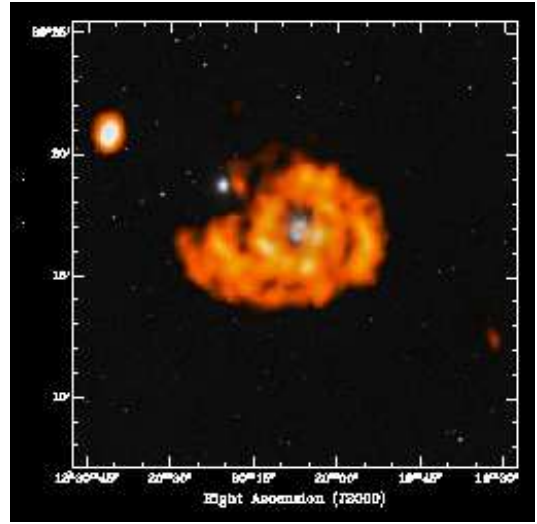,width=7cm,angle=-90}}
\caption{WSRT total \HI\ intensity (orange) of the galaxy NGC~4278 superimposed to
the optical image (grey).}
\end{figure}

\begin{figure*}
\centerline{\psfig{figure=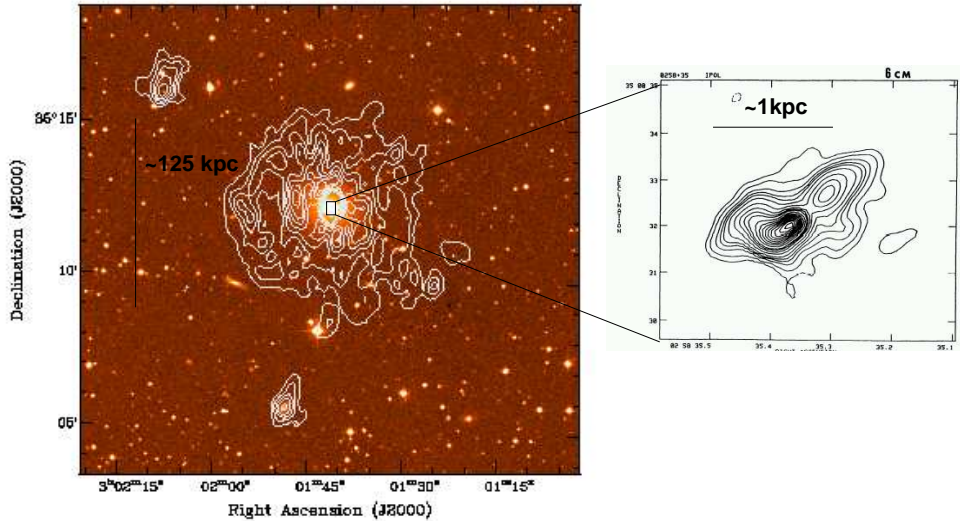,angle=0,width=13cm}
}
\caption{{\sl Left} \HI\ total intensity (contours) superimposed to an optical
image of B2~0258+35 (from Emonts et al. in preparation). {\sl Right} The
continuum image (from Fanti et al. 1986).}
\end{figure*}

\subsection{\HI\ in nearby Radio Galaxies}

The results presented above have been obtained for samples of ``normal''
early-type galaxies, therefore not biased toward radio-loud objects.  A study of
\HI\ in radio galaxies has been instead carried out by Emonts et al. (in
prep).  The sample includes more than 20 objects (both in the northern
and in the southern hemisphere). The selection and observations of the
northern sample are described in Emonts et al. (these proceedings).
The \HI\ mass limit of this survey is somewhat in between the surveys
described above (few time $10^7$ up to $10^8$ M$_\odot$) and therefore the
comparison between them is difficult.  Among the radio galaxies, 25\% of the
objects are detected in \HI\ in emission.  Interestingly, the most \HI-rich
objects (again with \HI\ masses up to $10^{10}$ $M_{\odot}$ and log M$_{\rm
\HI}$/L$_B$ is between -1 and 0) have the neutral hydrogen distributed in very
large disks with regular kinematics as found in the shallow survey described
in Sec. 2.1. An example of an extended disk of \HI\ around a radio galaxy is
shown in Fig.~5. Thus, as for those galaxies, the origin of the large amount
of \HI\ is likely to be major mergers of gas-rich disk galaxies.  The
formation of regular disk structures is explained in numerical simulations as
result of merger of similar size galaxies (from 1:1 to 1:4) with high angular
momentum (Barnes 2002; Burkert \& Naab 2003).  The gas from the progenitors
falls back at a late stage of the merger (after the starburst phase) and
settles in a disk.  The combination of \HI\ and study of the stellar
population (in progress) allows to put the AGN activity in the evolutionary
sequence of early-type galaxies (see the case of B2 0648+27 in Emonts et
al. these Proceedings).

Other radio galaxies appear to have a much smaller amount of \HI\ (with \HI\
masses between $10^{8}$ and $10^{9}$ $M_{\odot}$ and log M$_{\rm \HI}$/L$_B$
is between -2 and -1) and the neutral hydrogen is either distributed in disks
or in blobs or tails.  These objects appears to be more similar to what found
in the deep SAURON survey (see Sec. 2.2).  Thus, although the comparison is not
completely fair and based on small numbers statistic, we do not find so far
major differences in the \HI\ characteristics (detection rate, morphologies,
masses, etc.) between ``normal'' early-type and radio galaxies.  This may
indicate that indeed the radio-loud phase is just a short period in the life
of many (all?) early-type galaxies.  However, it should be noticed that in our
sample of radio galaxies all the large \HI\ disks have been detected so far
around compact radio sources, see Fig. 5 for an example. The reason for this
is not yet clear. Some possible explanations are given in Emonts et al. (these
Proceedings).

Finally, we would like to point out possible similarities between the large
quiescent Ly$\alpha$ structures detected in high-$z$ galaxies (Villar-Mart\'in
et al. 2002, 2005 in these proceedings) and the large \HI\
disks detected in low-$z$ radio galaxies. 



\begin{figure*}
\centerline{\psfig{figure=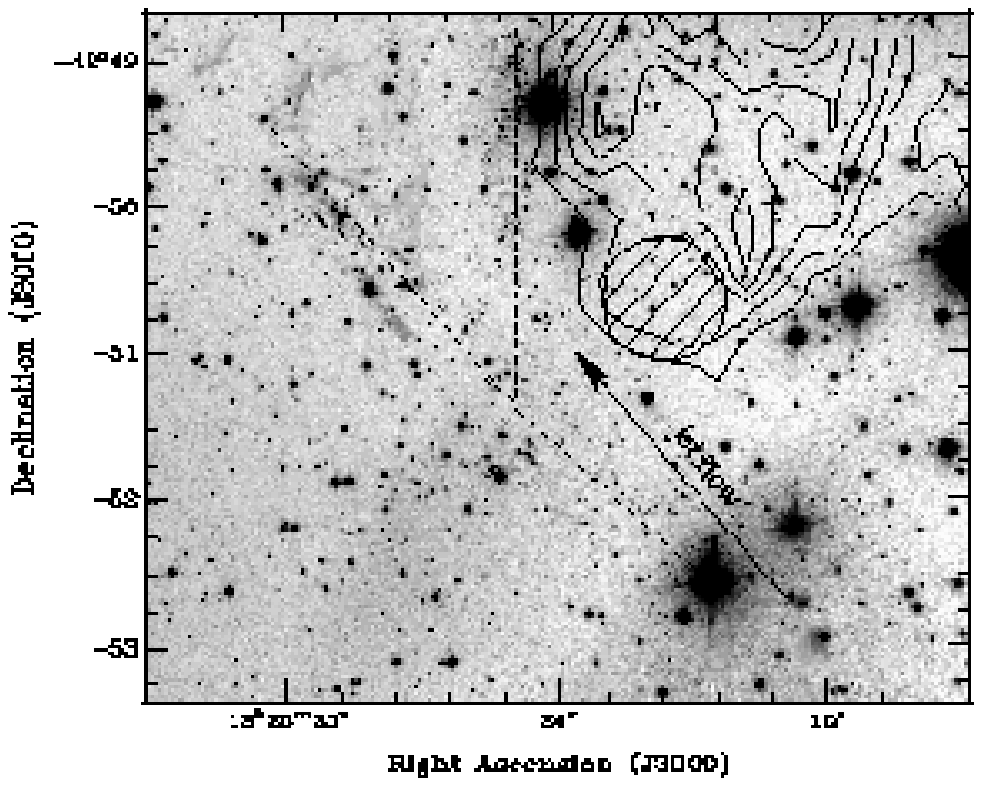,width=8cm,angle=-90}
\psfig{figure=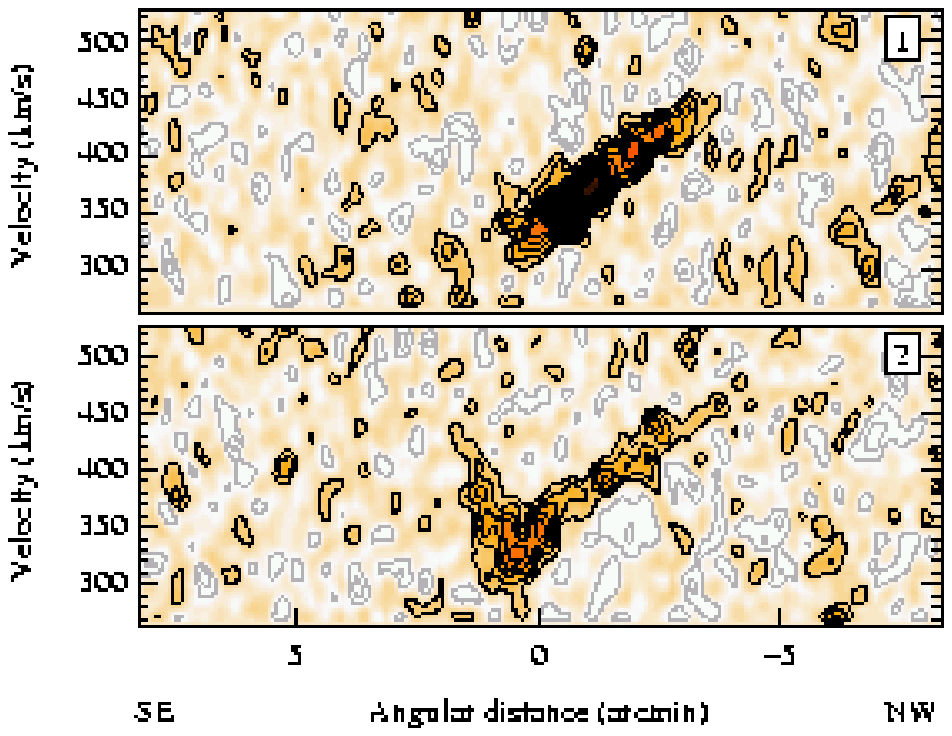,width=8cm,angle=-90}}
\caption{{\sl Left}  \HI\ contours of the southern region of the \HI\ cloud,
drawn on top of a broad band optical image (kindly provided by M.\
Rejkuba). The hatched area indicates the location where the anomalous \HI\
velocities are detected while the arrow indicates the location and the flow
direction of the radio jet. The filament of ionised gas is visible in the top
left.  The dashed lines roughly indicate the locations of young stars.  {\sl
Right} Position-velocity plots taken through the location of the anomalous
\HI\ (bottom) and, for comparison, taken just above that region (top). Taken
from Oosterloo \& Morganti (2005).}
\end{figure*}

\section{Jet-induced star formation}

As shown above, the host galaxies of radio-loud AGN can be gas rich. Thus, the
interaction between the non-thermal plasma ejected from the active nucleus and
the ISM of a galaxy can have important consequences and can be responsible for a
variety of phenomena in radio galaxies such as ionisation of the gas, AGN
driven outflows and jet-induced star formation. Such interactions are
considered to be particularly relevant in high redshift radio galaxies, as
they are typically living in a gas-rich environments (see e.g.\ van Breugel
2000 and references therein). As mentioned above, one aspect of jet-ISM
interaction is that it can trigger star formation and this is considered a
possible mechanism to explain the UV continuum emission observed in the host
galaxies of distant radio sources and the "alignment effect" between the radio
emission and this continuum (Rees 1989).  Detecting and studying star
formation produced by this mechanism in high-$z$ radio galaxies is very
challenging.  The only case where this has been done is 4C~41.17 (Dey et al.\
1997).  Because of the observational problems for high-redshift sources, it is
important to find nearby examples of star formation triggered by the radio jet
that can be studied in more detail.  The nearby, best examples are Centaurus A
(Oosterloo \& Morganti 2005) and the Minkowki's Object (van Breugel et al. 1985).

\subsection{Jet-induced star formation in Centaurus ~A}

In the case of Centaurus~A, new 21-cm \abHI\ observations of the large \abHI\
filament located about 15 kpc NE from the centre of this galaxy and discovered
by Schiminovich et al.\ (1994) have been carried out using the ATCA (Oosterloo
\& Morganti 2005).  This \abHI\ cloud is situated (in projection) near the
radio jet of Centaurus A (see Fig. 6), as well as near a large filament of
ionised gas of high excitation and turbulent velocities and near regions with
young stars.  The higher velocity- and spatial-resolution of the new data
reveal that, apart from the smooth velocity gradient corresponding to the
overall rotation of the cloud around Centaurus A (Schiminovich et al. 1994),
\abHI\ with anomalous velocities up to 130\kms\ is present at the southern tip
of this cloud (see Fig. 6 and Oosterloo \& Morganti 2005).  This is
interpreted as evidence for an ongoing interaction between the radio jet and
the \abHI\ cloud.  Gas stripped from the \abHI\ cloud gives rise to the large
filament of ionised gas and the cooling of the gas is then responsible for the
star formation regions that are found downstream from the location of the
interaction.
 From the displacement of the
young stars from the location of the anomalous velocities we derive a flow
velocity of about 100\kms,  very similar to the observed anomalous \abHI\ velocities.. 
The jet induced star formation appears to be fairly
inefficient, of the order of few percent.   
Recent numerical simulation have shown that radio jets can indeed drive
radiative shocks in interstellar clouds, causing them to compress and break up
into numerous dense, cloud fragments. These fragments survive for many
dynamical timescales and are presumably precursors to star formation (Mellema
et al. 2002; Fragile et al. 2004).

\begin{figure*}
\centerline{\psfig{figure=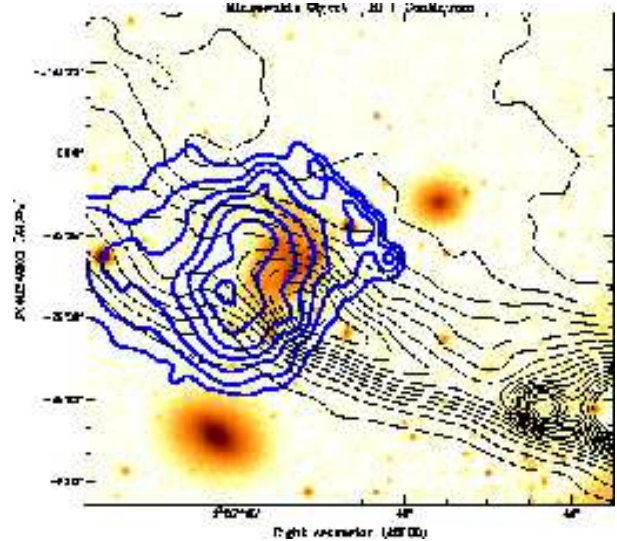,width=8cm,angle=-90}}
\caption{C-array \HI\ cloud (thick, blue contours) at the end of the radio continuum
jet (thin, black contours) superimposed onto a Keck image of the Minkowski's
Object (van Breugel et al. in prep). }
\end{figure*}

\subsection{Jet-induced Star Formation in ``Minkowski's Object''}

Minkowski's Object is a peculiar starburst galaxy (at $z$ = 0.0187) at end of
the jet from NGC~541 (van Breugel et al 1985; van Breugel et al. 2004; Croft
et al. 2004).  This object is an ideal candidate for a detailed study of the
effects of the interaction between the radio plasma and the ISM. Its
morphology is strongly suggestive of a collision of a low luminosity FR-I type
jet from NGC 541 with a gas rich cloud/object (see Fig.~7 and van Breugel et
al. 1985).

Recent C-array VLA observations have shown that a cloud of \HI\ is detected
just down-stream from the main star formation region (see Fig.~7).  This cloud
has a total \HI\ mass of $\sim 3 \times 10^8$ M$_\odot$ (corresponding to a
M$_{\rm \HI}$/ L$_B \sim 0.17$), it has the same transverse size as the radio
jet and consists of two main components separated at the centre of the radio
jet. A velocity gradient in the \HI\ of $\sim 60$ \kms\ is observed.

Although, as in the case of Centaurus~A, \HI\ is detected close to the
location of the star formation and radio jet, it is not clear whether the same
scenario can explain the case of the Minkowski's Object.
In this object the radio jet itself may be the cause
of the formation of the \HI\ as predicted in the jet-triggered radiative
cooling model (Fragile et al. 2004). This model shows  that star formation, and the
\HI\ that preceeds this, can occur in relatively warm gas due to radiative
cooling triggered by the radio jet. However, until stronger constrains to this
model will be available from new, higher resolution \HI\ observations  the
possibility that the origin of the starburst is instead due to a 
pre-existing peculiar galaxy or a pre-existing cold gas (like in the case of
Centaurus~ A) cannot be ruled out.

\subsection{Results from jet-induced star formation}

The kinematic of the \HI\ supports the idea that at least in two nearby
objects (Centaurus ~A and Minkowski's Objects) radio jets can trigger star
formation.  In both cases, \HI\ is observed in regions where jet-induced star
formation has been claimed to be present. The kinematics of the neutral
hydrogen, its morphology and its relation to ionized gas give constraints on
the on-going process. The next step will be a more detailed comparison between
the results from the observations (in particular from high resolution \HI\
data) and the results from numerical simulations.  Because the gas densities
and the AGN activity are higher in early Universe, jet-induced star formation
is, therefore, likely to be a more common phenomenon in high-$z$ radio
galaxies.

\begin{figure}

\centerline{\psfig{figure=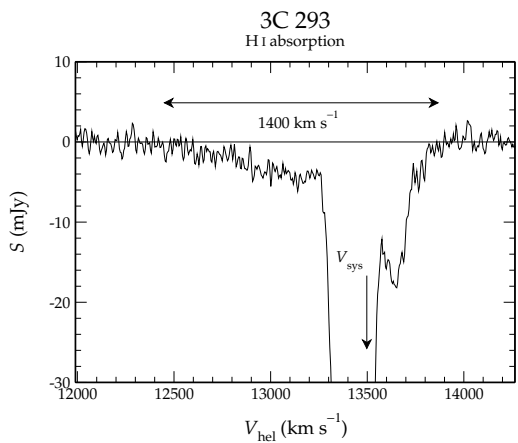,angle=0,width=7cm}}
\caption{ A zoom-in of the \ion{H}{i}\ absorption spectra of 3C~293 clearly
showing the broad \ion{H}{i}\ absorption.  The spectra are plotted in flux
(mJy) against optical heliocentric velocity in km/s. From Morganti et al. 2003.}
\end{figure}

\section{Fast \HI\ outflows}

Gaseous outflows appear to be a widespread phenomenon in galaxies, both in the
local as well as in the far-away Universe (e.g. Crenshaw et al. 2003; Veilleux
et al. 2005; Frye et al. 2002). They can be driven
by super-winds in the starburst phase or by the energy released in the active
phase of the nucleus. AGN-driven outflows of ionized gas have been detected in
many nearby galaxies. These outflows are a key ingredient in  galaxy
evolution.  The correlation found between the mass of the super-massive
black-hole and the mass of the central bulge of the galaxy are most
easily understood in terms of feedback models (Silk \& Rees 1998). Numerical
simulations suggest that the energy released by a quasar expels enough gas to
quench both star formation and further black-hole growth (e.g.  di Matteo,
Springel, Hernquist 2005).

Recent results underline the importance of radio jets in producing such
outflows. These outflows have been found not only associated with ionized gas
(see Tadhunter these Proceedings) but also with neutral hydrogen.  Recent
sensitive, broad-band 21-cm observations of the radio sources IC~5063
(Oosterloo et al. 2000) and 3C~293 (see Fig. 8; Morganti et al. 2003; Emonts
et al. 2005) have revealed that fast outflows of neutral hydrogen can occur in
galaxies with an AGN. In addition to these two initial objects, more cases of
broad and blueshifted \HI\ absorption have been found using very sensitive and
broad-band 21-cm radio observations with the Westerbork Synthesis Radio
Telescope. Thus, fast and massive outflows may be common, in particular in
young powerful radio sources (Morganti et al. 2005b). Interestingly, the mass
outflow rates detected for the neutral gas are much larger than those
typically found for the ionised gas and are large enough to have a significant
impact on the evolution of the galaxies.  The WSRT broad band covers $\pm
2000$ \kms\ around the central velocity (systemic velocity of the galaxy). The
observed borad \HI\ absorption features have a width up to 2000 \kms\ and a
typical optical depth $<<1$\%, corresponding to a column density few times
$10^{21}$ cm$^{-2}$ (for T$_{spin}$ = 1000 K) The detected \HI\ absorption
features are mostly blueshifted therefore corresponding to gas outflows. In
order to understand the origin of such fast outflows it is important to know
the location where the outflow is occurring.

\begin{figure}
\centerline{\psfig{figure=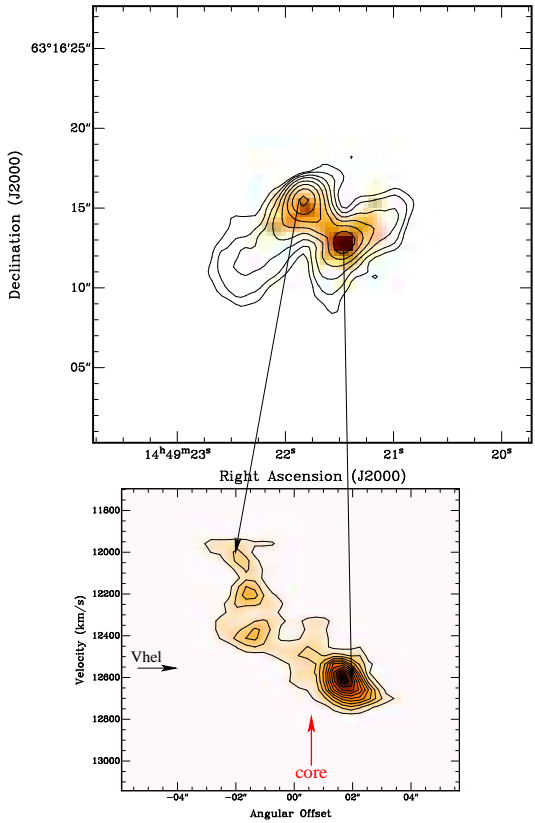,angle=0,width=7cm}}
\caption{Panel showing {\sl (top)} the radio continuum  image
of 3C~305.  {\sl (Bottom)}
  The position-velocity plot from a slice passing through the two
  lobes. The broad \HI\ absorption is detected against the NE radio lobe,
about 1.6 kpc from the nucleus. The contour levels for the continuum image are 10
  \mJybeam\ to 830 \mJybeam\ in  steps of a factor 2.  The
  grey scale image represents the total intensity of the \HI\
  absorption. The contour levels of the \HI\ are $-0.7$, ..., -7.7
  \mJybeam\ in  steps of 0.7 \mJybeam.  The arrow represents the
  systemic velocity. Taken from Morganti et al. (2005c)}
\end{figure}

\begin{figure}
\centerline{\psfig{figure=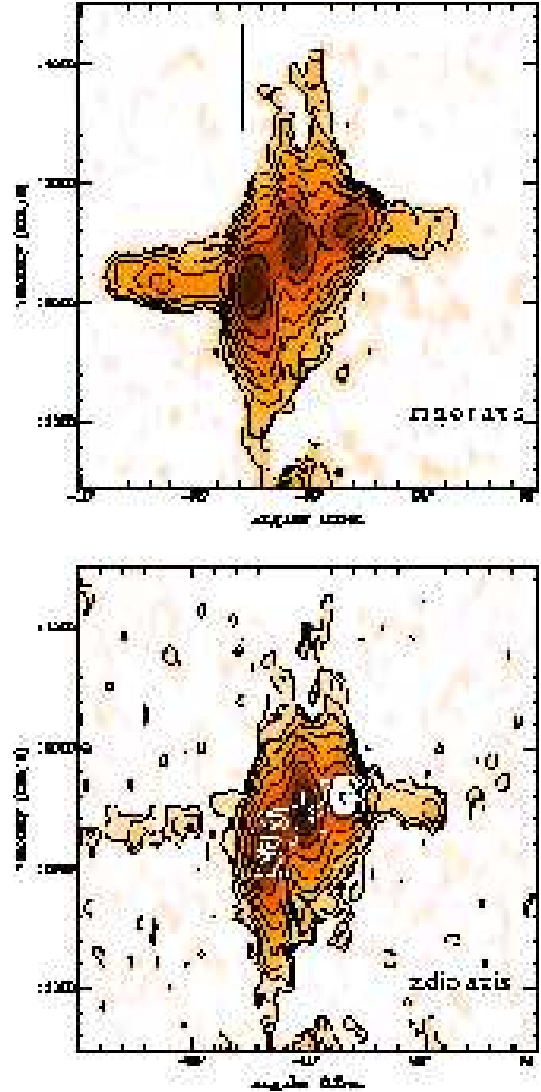,angle=0,width=7cm}}
\caption{{\sl Top} WHT spectrum of the \OIII\ region of 3C~305 (after the
subtraction of the continuum from the galaxy) taken in p.a.\ 60$^\circ$, i.e.\
along the galaxy's major axis (NE to the left, SW to the right).  The two
arrows represent the approximate position of the peak of the radio lobes.
{\sl Bottom} WHT spectrum of the \OIII\ region of 3C~305 (black contours and
grey scale) taken in p.a.\ 42$^\circ$, i.e. along the radio axis.  White
contours represent the \HI\ position-velocity plot taken along the radio axis
(as in Fig.\ 9).}
\end{figure}

\subsection{The case of 3C~305}

High-spatial resolution 21-cm \abHI\ VLA observations were obtained for the
radio galaxy 3C~305 (Morganti et al. 2005c).  These new high-resolution data
show that the $\sim 1000$ \kms\ broad \HI\ absorption, earlier detected in
low-resolution WSRT observations, is occurring against the bright, eastern
radio lobe, about 1.6 kpc from the nucleus (see Fig.~9).  

We also used new optical spectra taken with the WHT to make a detailed
comparison of the kinematics of the neutral hydrogen with that of the ionised
gas (see Fig. 10).  The striking similarity between the complex kinematics of
the two gas phases suggests that both the ionised gas and the neutral gas are
part of the same outflow.  Earlier studies of the ionised gas (Heckman et
al. 1982; Jackson et al.  2003) had already found evidence for a strong
interaction between the radio jet and the ISM at the location of the eastern
radio lobe. These results show that the fast outflow produced by this
interaction also contains a component of neutral atomic hydrogen. The most
likely interpretation is that the radio jet ionises the ISM and accelerates it
to the high outflow velocities observed. These observations demonstrate that,
following this strong jet-cloud interaction, not all gas clouds are destroyed
and that part of the gas can cool and become neutral.  The mass outflow rate
measured in 3C~305 (but also in other objects, see below) is comparable,
although at the lower end of the distribution, to that found in Ultra Luminous
IR galaxies.  This suggests that AGN-driven outflows, and in particular
jet-driven outflows, can have a similar impact on the evolution of a galaxy as
starburst-driven superwinds.

\subsection{Results from fast \HI\ outflows}

The results obtained so far show that fast outflows of neutral hydrogen can be
produced by the interaction between the radio jet and the surrounding dense
medium. The presence of neutral gas in these regions indicates that the gas
can cool very efficiently following a strong jet-cloud interaction.
Interestingly, the associated mass outflow rates range from a few tens to
about almost hundred M$_\odot$ yr$^{-1}$, comparable to (although at the lower
end of the distribution) the outflow rates found for starburst-driven
superwinds in Ultra Luminous IR Galaxies (Rupke et al., 2002; Heckman
2002). Thus, as these superwinds, the massive, jet-driven HI outflows in the
radio-loud AGN can have a major impact on the evolution of the host galaxy.
High-$z$ \HI\ absorbers have been also found in Ly$\alpha$ profiles (although
with lower column density). As for the low-$z$ cases, a possible way to
explain these absorptions is via highly supersonic jet expanding into the
dense medium of a young radio galaxy that then will be surrounded by an
advancing quasi-spherical bow shock (Wilman et al.  2003; Krause 2002).  Thus,
the detailed results obtained for the nearby objects can help in understanding
better the mechanism at work in their far away cousins.

\section{Conclusions}

This review aimed to illustrate the importance that studies of the neutral
hydrogen can have for the understanding of different phenomena observed in
radio galaxies. In nearby radio galaxies, the \HI\ is beginning to tell us
about the origin of the host galaxy and about the presence and location of
fast gaseous outflows, that appear to have a significant impact on the
evolution of the galaxy.  Nevertheless, despite the recent progress, a number
of questions remain still open.  For example: what is the link (if any)
between large \HI\ disks (low-$z$) and quiescent Ly$\alpha$ structures
(high-$z$)? what are the details of the physical process that produces fast
\HI\ outflows? How common/important is the jet-induced starformation at
high-$z$ (and at low-$z$)?

Unfortunately at present deep studies of the neutral hydrogen as presented in
this review are only limited to nearby galaxies. The need for the new
generation of radio telescopes, and in particular for the Square Kilometer
Array is clear. A detailed overview of what this new instrument will do for us
can be found in the ``SKA Science Case'' (Carilli \& Rawlings 2004).
Hopefully we do not only have to wait for SKA but we may have some answers to
these questions before the next meeting!

\acknowledgements

I would like to thanks the organizing commetee and in particular Montse
Villar-Martin for organizing and inviting me at this very pleasant and
interesting workshop.  The results presented in this review would not have
been obtained without the help of my collaborators. In particular I would like
to thanks T. Oosterloo, C.N. Tadhunter, B. Emonts, E. Sadler, T. de Zeeuw, W. van
Breugel, J. van Gorkom.


\end{document}